\documentclass[a4paper]{PoS}

\graphicspath{{figs/}}
\usepackage{style}
\newcommand{\FIG}[1]{Fig.~\ref{fig:#1}}
\newcommand{\EQ}[1]{Eq.~\eqref{#1}}
\newcommand{\SEC}[1]{Sec.~\ref{sec:#1}}

\title{NPR determination of quark masses from the HISQ action}

\ShortTitle{NPR determination of quark masses from the HISQ action}

\author{\speaker{Andrew T.\ Lytle} \\ 
        SUPA, School of Physics and Astronomy \\
        University of Glasgow, Glasgow, G12 8QQ, UK \\
        E-mail: \email{Andrew.Lytle@glasgow.ac.uk}}

\author{HPQCD Collaboration}

\abstract{
I report on a calculation of bilinear Z-factors needed for determining
$Z_m$ using non-perturbative renormalization (NPR) on $n_f=2+1+1$
HISQ ensembles.  RI/MOM and RI/SMOM schemes are studied.
These will provide an independent determination of
quark masses in addition to other methods being used by the HPQCD
collaboration.
}

\FullConference{The 33rd International Symposium on Lattice Field Theory\\
		14 -18 July 2015\\
		Kobe International Conference Center, Kobe, Japan}

\begin{document}

\section{Introduction}
Knowledge of renormalized quark mass parameters has 
been improved significantly
in recent years from lattice QCD methods.  Precise determination of these are
important for BSM phenomenology and the search for new physics.
For example,
precision measurements of Higgs couplings to $b$ and $c$ quarks
at future colliders
must be combined with similarly precise SM mass determinations
in order to detect deviations from the SM~\cite{Lepage:2014fla}. 
Currently, uncertainties in $b$ and $c$ masses
are at the percent to few-percent level.  Reducing uncertainties to
the sub-percent level 
is a challenge which requires input from multiple
lattice groups, as well determinations from multiple methods. 
Both of 
these help ensure a robust estimate of uncertainties which may stem from
systematics in either formulation or method.

There have been several recent lattice determinations of charm mass from
different groups, which are broadly in agreement with one another
(a recent summary of these can be found in~\cite{Lytle:2015oja}).
The most precise of these at present come from 
the ``current-current correlator'' 
method~\cite{Allison:2008xk,Chakraborty:2014aca} in simulations using a
Highly Improved Staggered Quark (HISQ) action, which allows a relativistic
treatment of charm, on $n_f=2+1$ and $n_f=2+1+1$ ensembles,
plus continuum QCD perturbation theory through $\O(\alpha_s^3)$.
Results from an application of this method using domain-wall fermions
were also presented at this conference~\cite{Nakayama:2015}.
A precise value for $m_c$ can be translated
into precise $m_s, m_{ud}$ values using bare
quark mass ratios~\cite{Davies:2009ih}. 

The $c,s,ud$ masses can also be obtained from $n_f=2+1+1$ HISQ 
simulations using non-perturbative renormalization (NPR) methods.
RI/MOM results on $n_f=2+1$ asqtad ensembles were presented 
in~\cite{Lytle:2009xm}, 
and in an ongoing determination with HYP-smeared
valence quarks~\cite{Kim:2013bta,Jeong:2015omj}.
Here we present first results for both RI/MOM and RI/SMOM schemes
using the HISQ action.
\SEC{npr} describes the basic methodology with Secs.~\ref{sec:npr-mom} 
and~\ref{sec:npr-smom} presenting more details on the
RI/MOM and RI/SMOM schemes respectively.

\section{NPR method} \label{sec:npr}
The $\MSbar$ renormalized quark mass is related to the bare lattice input mass
$m_0$,
\begin{equation}
m^{\MSbar}(\mu) = Z_m^{\MSbar}(\mu, 1/a) \, m_0 \,.
\end{equation}
In practice it is difficult to compute the factor 
$Z_m^{\MSbar}(\mu, 1/a)$ beyond $\O(\alpha_s)$ using lattice perturbation
theory, especially with improved actions 
($Z_m$ was calculated
to $\O(\alpha_s^2)$ using the asqtad action in~\cite{Mason:2005bj}).

An alternative method was proposed in~\cite{Martinelli:1994ty}, 
which breaks the problem into two steps.  
The first step makes use of an intermediate
Regularization Independent (RI) scheme that is well-defined
both on the lattice and in the continuum.
Using this scheme the renormalization conditions are imposed directly
on lattice correlation functions.  In this way one obtains $Z$-factors
to all orders in $\alpha_s$,  however this determination
will be sensitive to discretization errors as well as non-perturbative
effects.  Therefore the method requires that the renormalization condition
be applied within a range
\begin{equation} \label{npr-window}
\Lambda_{\text{QCD}} \ll |p| \ll \frac{\pi}{a} \,.
\end{equation}
This ensures both discretization effects and non-perturbative contributions
are small.

The second step then uses a conversion factor, computed using continuum
perturbation theory, to convert the $Z$-factor to the $\MSbar$ scheme.
The continuum conversion factors are generally known to higher order
than the lattice to continuum factors. 
It should be noted results obtained from the first step
are universal and the second step is only necessary if one requires
results in a non-RI scheme such as $\MSbar$.

\subsection{Calculation details}
Results presented here were calculated on a single
coarse ($a \approx 0.12$ fm) $n_f=2+1+1$ HISQ ensemble~\cite{Bazavov:2010ru},
with lattice volume $L^3 \times T$, $L=20$, $T=64$, and
with sea masses of $am_c=0.635$, $am_s=0.0509$, and $am_l = 0.0102$.
The configurations are fixed to Landau gauge.
Propagator inversions were with valence masses
$am_{\text{val}} = 0.0509$, 0.0102, and 0.00501.

Inversions were done using momentum sources, and with a variety of momentum
values.  For the RI/MOM results the momenta either have the form
$2 \pi (x/L,\, x/L,\, x/L,\, 3x/T)$ for $x=1,2,3$ or $(x, x, x, 0)$. 
The latter were computed using
twisted boundary conditions and are not restricted to lattice Fourier
modes. RI/SMOM results have $p_1 = 2\pi (x/L,\, 0,\, x/L,\, 0)$ and
$p_2 = 2\pi(x/L,\, -x/L,\, 0,\, 0)$ with $x=3,4,5$.

\subsection{RI/MOM scheme} \label{sec:npr-mom}
We calculate off-shell Landau gauge-fixed 
Green functions of bilinear operators with
external quark states, these have the form
\begin{equation} \label{Gij}
G^{ij}_\G(p) = \langle 
q^{i}(p) \( \sum_x \bar{q}(x) \G q(x) \) \bar{q}^{j}(-p)
\rangle_{\text{amp}} \,.
\end{equation}
The $i$ and $j$ indices represent both spin and color.
The renormalization factors are obtained by requiring that
an appropriate trace of the correlation function in the interacting
theory equal its tree-level value.
\begin{equation}
\Lambda_\G(p) \equiv  \frac{1}{12} \Tr \[ \G \, G_\G(p)\] \simeq
\frac{Z_q(p)}{Z_\G(p)}
\end{equation}

The RI and $\MSbar$ schemes satisfy $Z_m = Z_S^{-1} = Z_P^{-1}$,
so that with the wavefunction renormalization factor $Z_q$
it is possible to obtain $Z_m$ from the scalar and pseudoscalar
correlators.  $Z_q$ may be obtained from the momentum-space (polespace)
propagator itself.
\begin{equation}
Z'_q(p'^2) = - \frac{i}{12 N_T} 
\sum_\mu \frac{p'_\mu}{p'^2} \ps{\gamma_\mu}{1} S^{-1}(p')
\end{equation}
(The polespace propagator assumes a continuum-like form but with $N_T=4$
taste degrees of freedom, details may be found in~\cite{Lytle:2013qoa}.)
In principle $Z_m$ could be obtained from the trace of the inverse
propagator, but this quantity also contains a quark condensate contribution
which is significant for momenta satisfying \EQ{npr-window}.
This condensate also strongly affects $\Lambda_P$, causing it to
differ from $\Lambda_S$ especially at low momentum where it
is not suppressed.

\begin{figure}[th]
\includegraphics[width=0.49\textwidth]{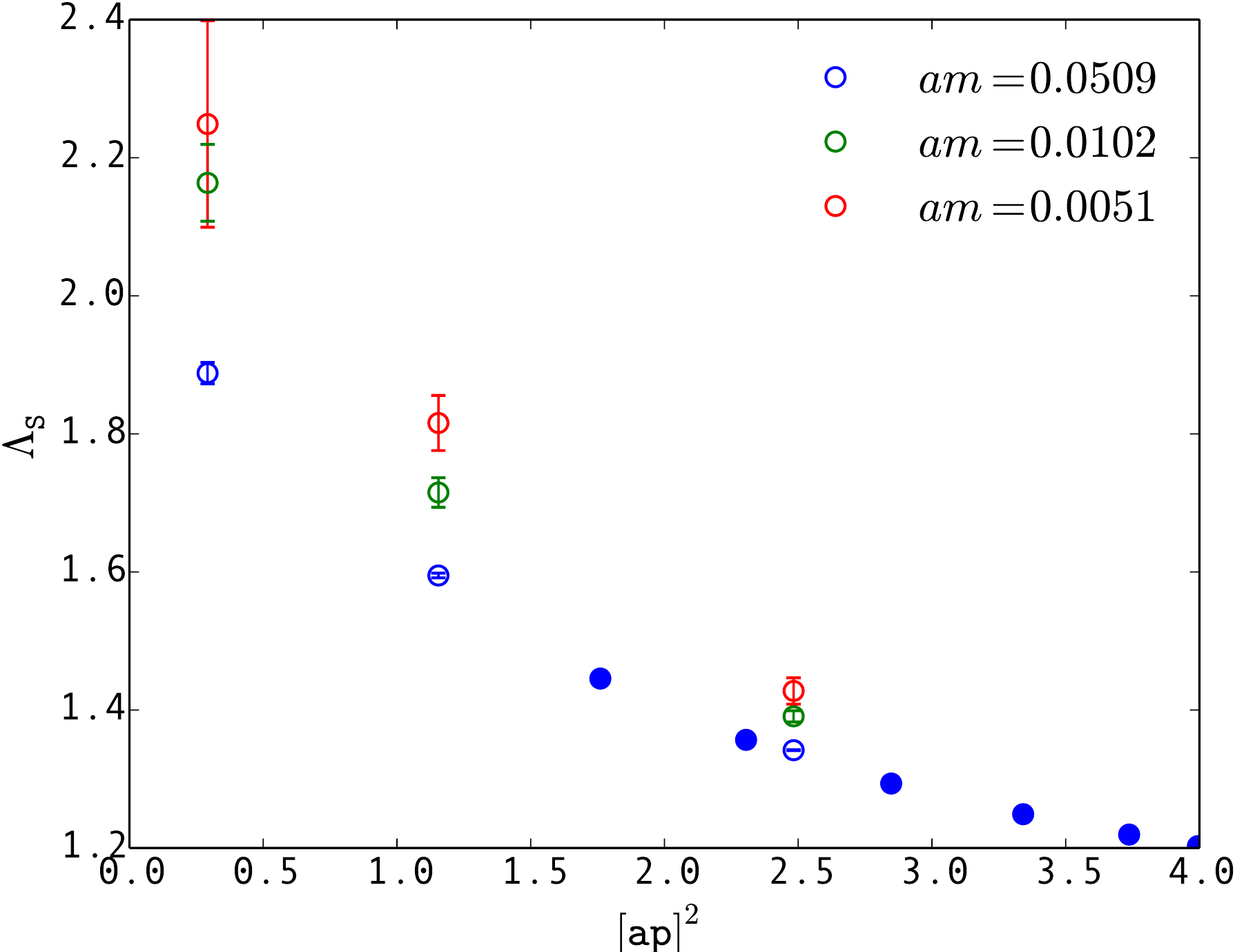}
\includegraphics[width=0.49\textwidth]{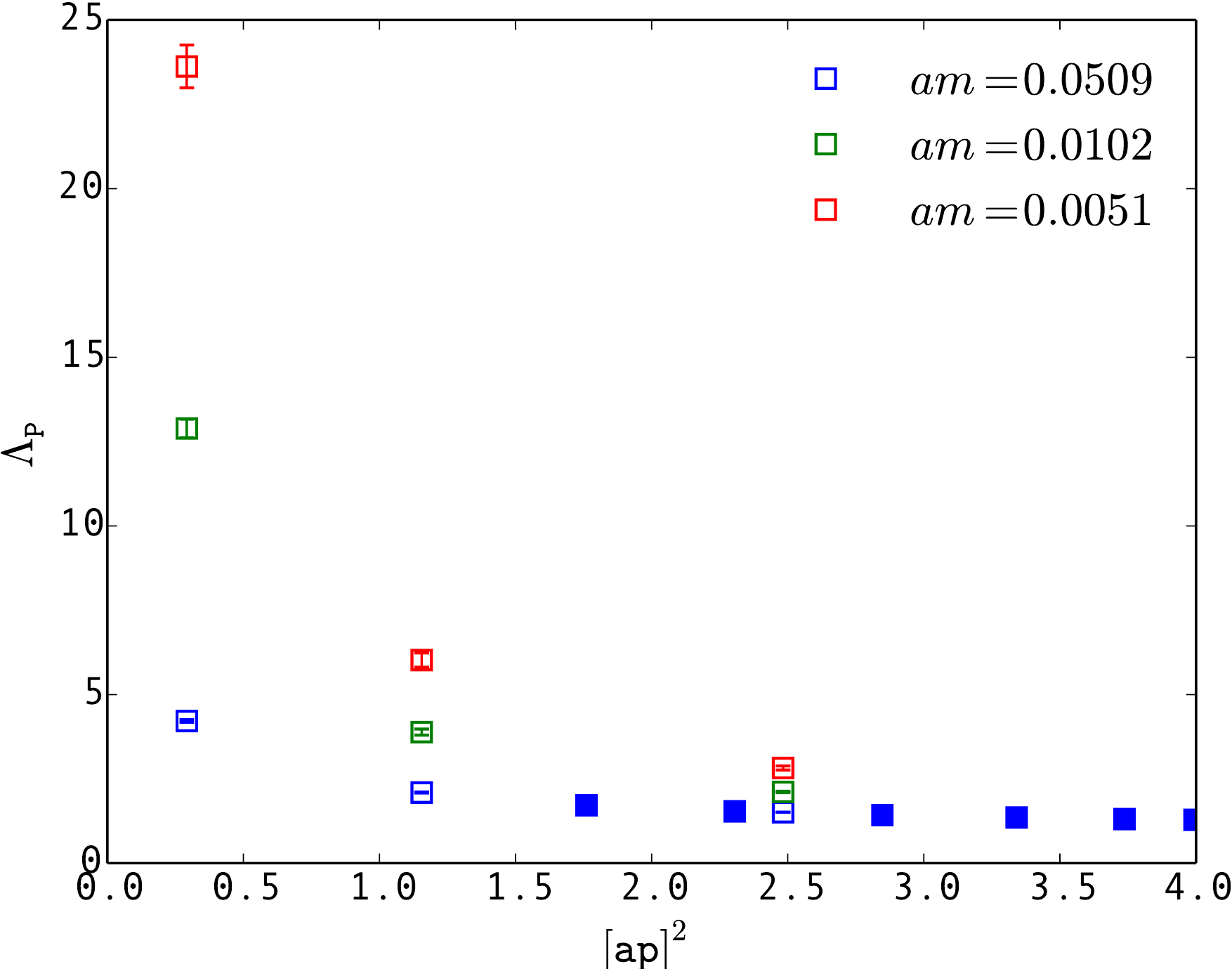}
\caption{
[Note different y-axis scales left vs.\ right] 
(Left) $\Lambda_S$ at three valence masses and for a range of momenta.
The filled points were obtained using twisted boundary conditions.
(Right) $\Lambda_P$ for the same masses and momenta.
\label{fig:LambdaS}}
\end{figure}

Some results for $\Lambda_S$ and $\Lambda_P$ are shown in 
\FIG{LambdaS}.  $\Lambda_P$ exhibits strong mass dependence due
to the condensate term, especially in the infrared, while $\Lambda_S$
has much milder behavior.  These two quantities are compared more
directly in \FIG{LambdaPS} (left), and it is evident that they
approach one another at large $p^2$ where the condensate term is suppressed.

\begin{figure}
\centering
\includegraphics[width=0.49\textwidth]{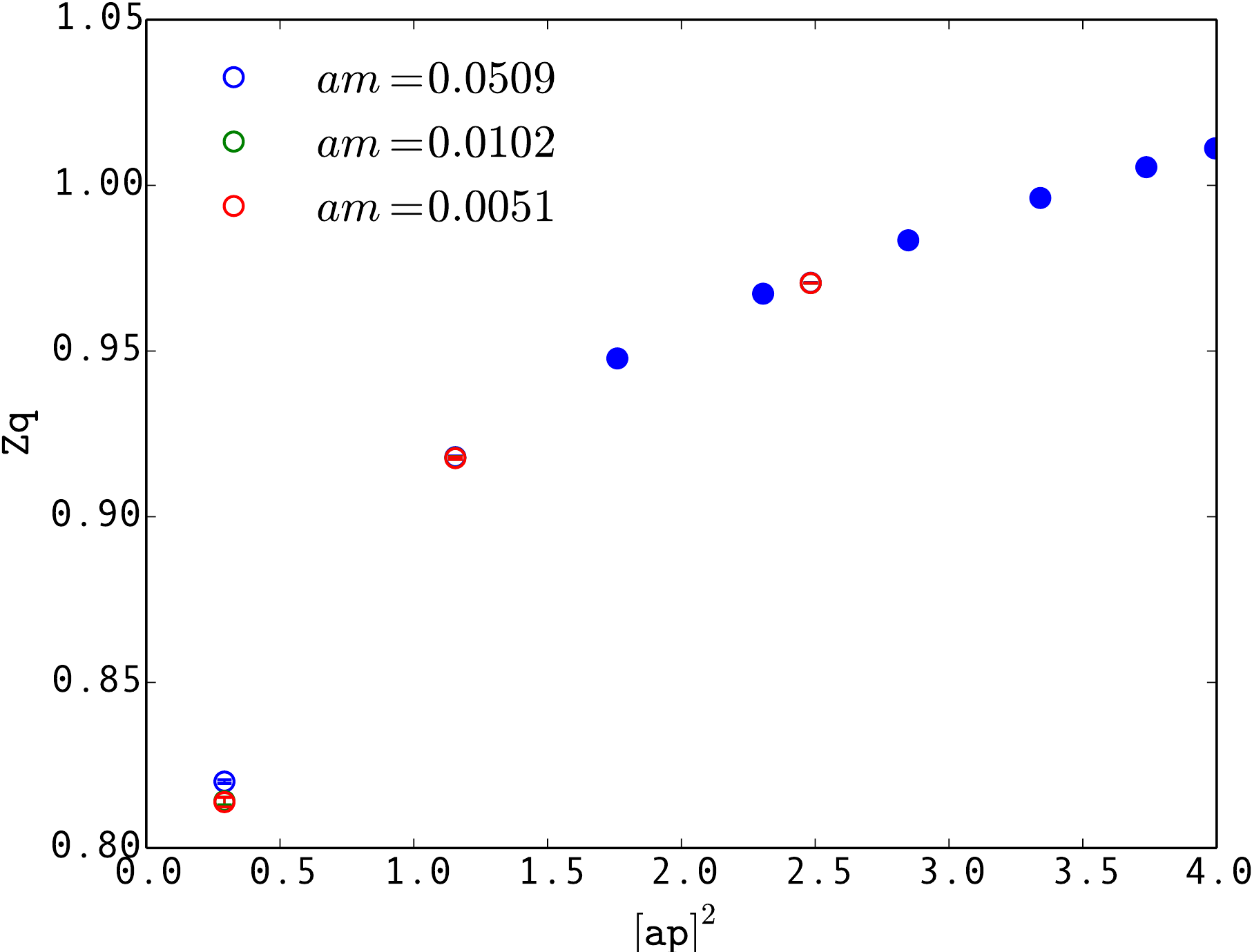}
\caption{$Z'_q$ extracted from the staggered polespace
propagator at three masses.  The filled points were obtained using
twisted boundary conditions.
\label{fig:Zq}}
\end{figure}

Results for $Z_q$ are shown in \FIG{Zq}.  This quantity has
almost no visible mass dependence.  The filled points were computed using
twisted boundary conditions, which allow for continuous variation of
$|p|$ keeping the orientation of $p$ fixed~\cite{Arthur:2010ht}.  
This is useful for
combining results from different lattice spacings and gives smooth
curves as a function of $(ap)^2$ because the lattice momenta all belong
to the same hypercubic representation.  These points have a different
momenta orientation from the others in \FIG{Zq} 
((x,x,x,0) vs.\ (x,x,x,3x)) and so give some indication of the size
of hypercubic artifacts.

\subsection{RI/SMOM scheme} \label{sec:npr-smom}
Due to infrared sensitivity exhibited by the MOM scheme,
it may be preferable to extract $Z_m$ using RI/SMOM intermediate
schemes first formulated in~\cite{Sturm:2009kb}.  
SMOM schemes have several advantages:
\begin{itemize}
\item Significantly reduced infrared sensitivity.
\item Reduction in mass dependence of $Z_{S,P}$ observed.
\item Matching factors to $\MSbar$ scheme closer to 1.
\end{itemize}

The SMOM scheme uses a different kinematic setup than the MOM scheme.
Whereas in the MOM scheme there is a single momentum $p$ 
(see \EQ{Gij}), the SMOM scheme uses separate momenta $p_1$, $p_2$
on each leg with $p_1 - p_2$ inserted at the vertex. Furthermore
these momenta satisfy the constraint $p_1^2 = p_2^2 = (p_1 - p_2)^2$
to maintain a single renormalization scale. The MOM and SMOM setups
are also referred to as ``exceptional'' and ``non-exceptional'' respectively.

\begin{figure}[th]
\includegraphics[width=0.49\textwidth]{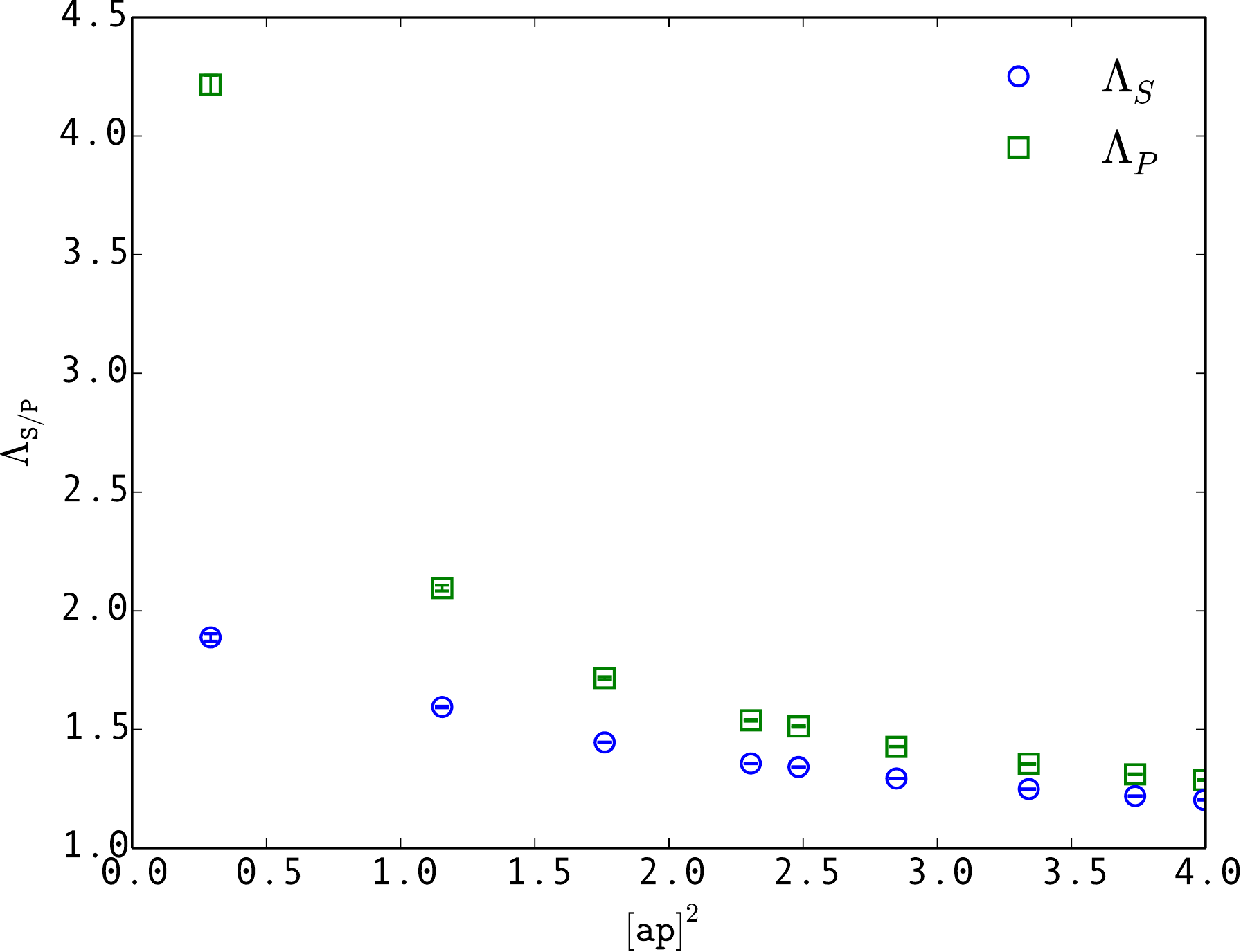}
\includegraphics[width=0.49\textwidth]{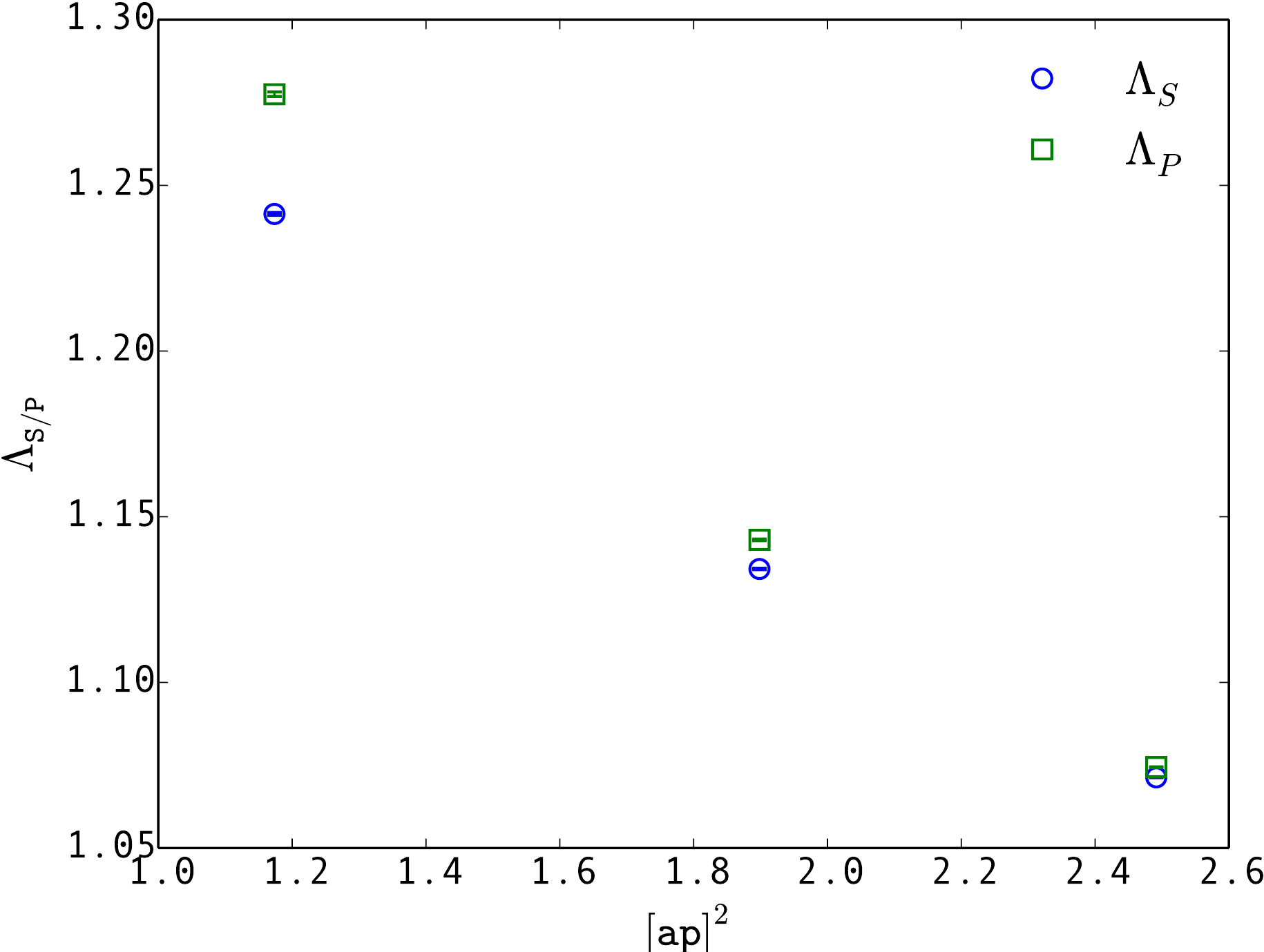}
\caption{
[Note different y-axis scales left vs.\ right]
(Left) Comparison of $\Lambda_S$ and $\Lambda_P$ as a function of
$(ap)^2$ for $am_{\text{val}}=0.0509$ using the exceptional scheme.
(Right) Same comparison but using the non-exceptional scheme. 
\label{fig:LambdaPS}}
\end{figure}

\begin{figure}[th]
\includegraphics[width=0.49\textwidth]{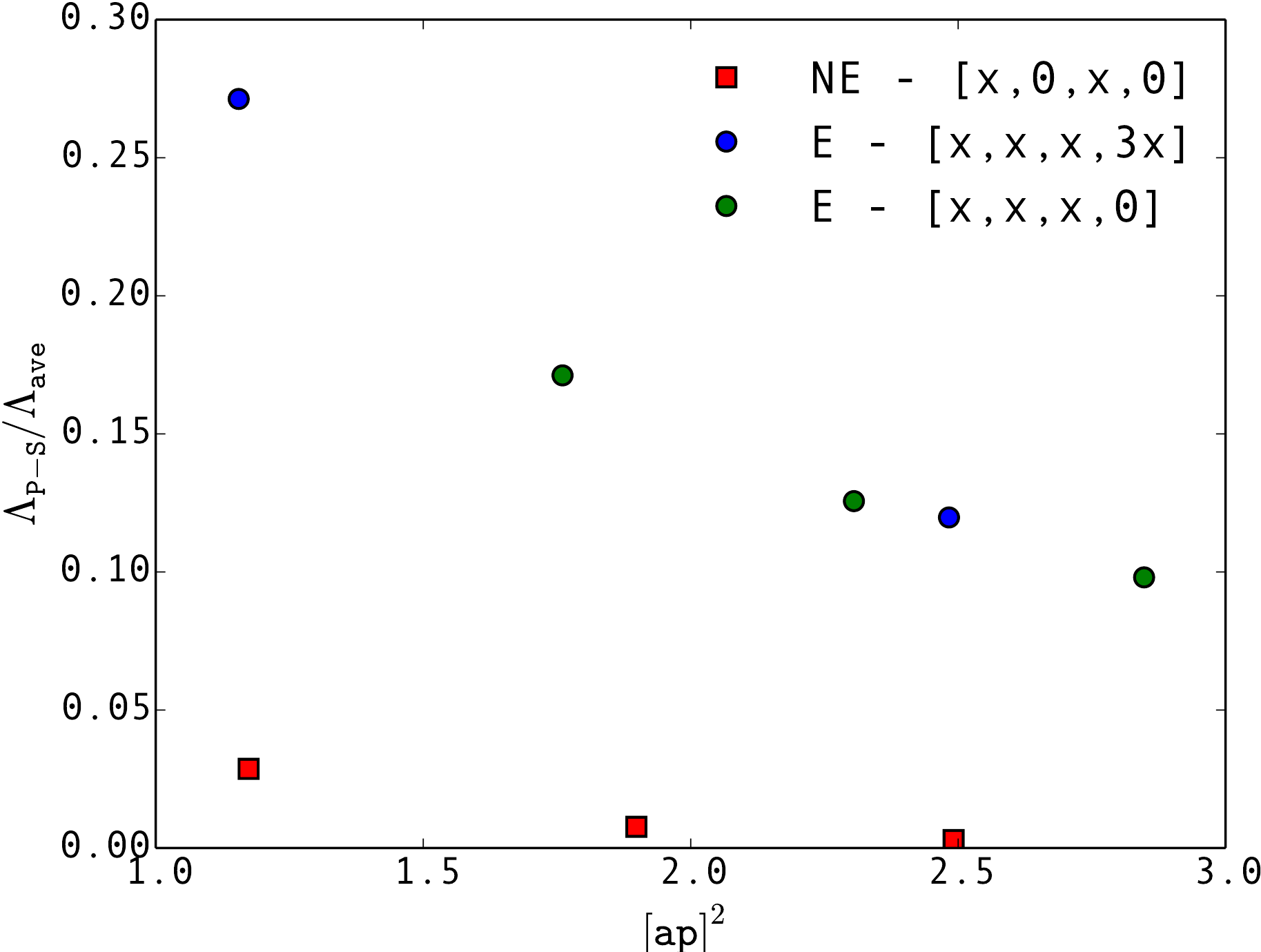}
\includegraphics[width=0.49\textwidth]{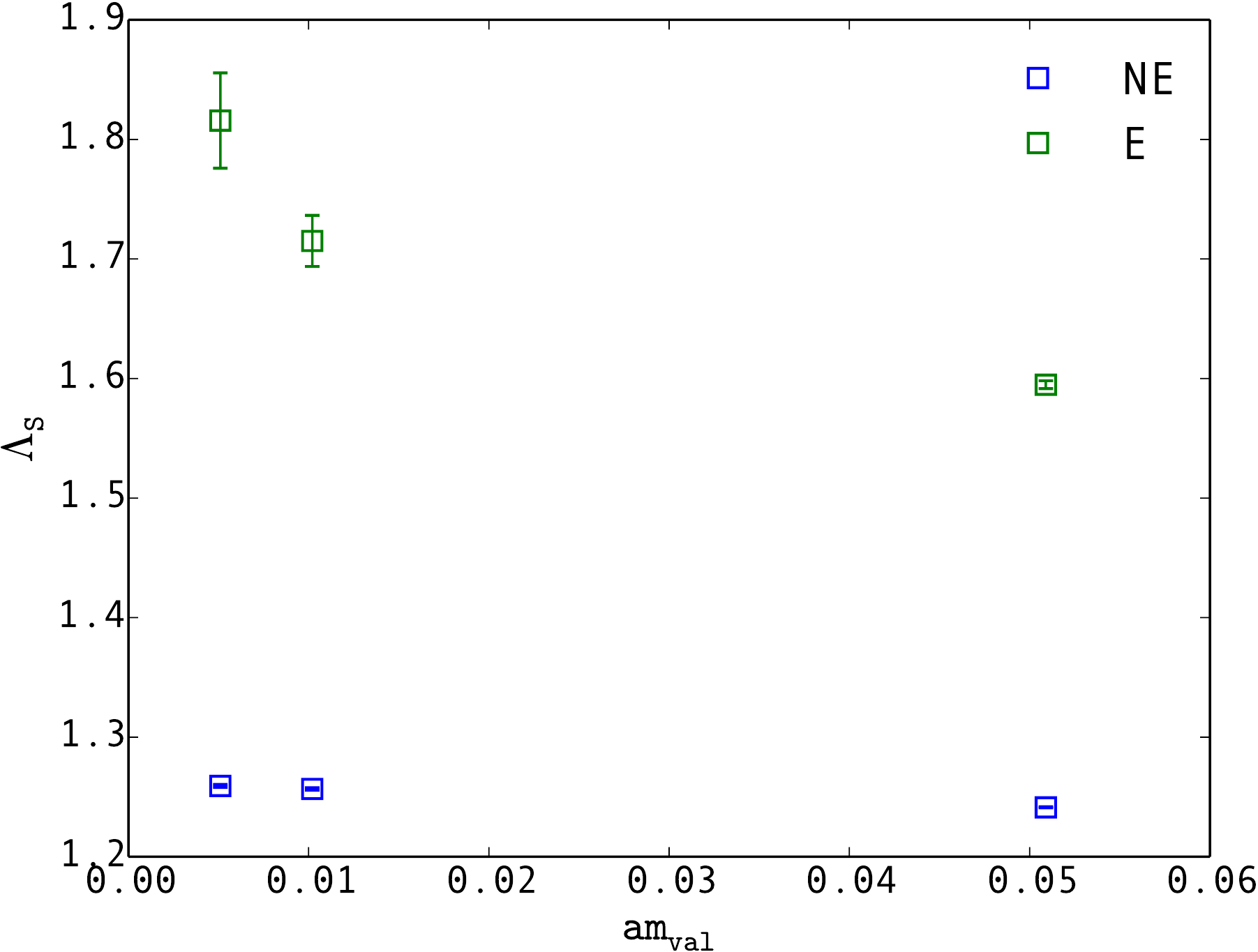}
\caption{
(Left) Difference of $\Lambda_P$ and $\Lambda_S$ divided by
their average for $am_{\text{val}}=0.0509$ 
using exceptional (blue and green circles) and non-exceptional
(red squares) kinematics. 
(Right) $\Lambda_S$ at fixed $(ap)^2 \approx 1.2$ as a function
of valence quark mass for the exceptional (green squares) and non-exceptional
(blue squares) schemes.
\label{fig:LambdaPmS}}
\end{figure}

Results for $\Lambda_S$ and $\Lambda_P$ obtained from the SMOM setup
at a single valence mass 
are shown in \FIG{LambdaPS} (right).
Now $\Lambda_P$ is much closer to $\Lambda_S$ over the full $(ap)^2$
range and especially at large $(ap)^2$ indicating effective suppression
of the condensate contribution.
This is shown further in \FIG{LambdaPmS} (left) which plots the
difference between $\Lambda_P$ and $\Lambda_S$ over their average
for both the exceptional and non-exceptional schemes.
Whereas the exceptional scheme shows an $\O(10\%)$ difference even at the
highest momenta studied the non-exceptional difference is at most
a few-percent and sub-1\% at large momentum.

\FIG{LambdaPmS} (right) compares the mass dependence of
$\Lambda_S$ at fixed momentum in the exceptional and non-exceptional
schemes.  
There is a rather strong mass dependence in the exceptional case.
For the non-exceptional scheme the variation is slight, from
valence mass at the physical strange down to $m_s/10$.
This is significant in practice because the lattice results are
obtained with non-zero masses and extrapolated to the chiral limit.

\section{Conclusions}
We have presented initial results for calculations of $Z_m$ using NPR
techniques on $n_f=2+1+1$ HISQ ensembles.  We studied both exceptional
(RI/MOM) and non-exceptional (RI/SMOM) schemes.  Results in the SMOM
scheme exhibit decreased sensitivity to the infrared, and a significantly
reduced dependence on valence mass is also observed, as compared to the
MOM scheme. These features, along with a $\MSbar$ 
matching factor which is close to 1, 
should prove useful in precision determination of $Z_m$.

The calculations presented here have been limited to a single, coarse
lattice ensemble.  In the future we will extend the work to 
fine ($a \approx 0.09$ fm) and
superfine ($a \approx 0.06$ fm) lattices.  It will also be important
to use ensembles with varying sea-quark masses for a given lattice spacing
in order to understand the approach to the chiral limit.

\section{Acknowledgements}
I would like to thank Christine Davies and Jonna Koponen for useful
discussions. Computations were carried out on the Darwin supercomputer,
part of STFC's DiRAC facility.

\end{document}